\documentstyle[12pt]{article}
\title{
From Effective BCS Action to Vortex Dynamics.}
\author{ G.E. Volovik\\
Low Temperature Laboratory\\
Helsinki University of Technology\\
Otakaari 3A, 02150 Espoo, Finland\\
and\\
L.D. Landau Institute for Theoretical Physics, \\
Kosygin Str. 2, 117940 Moscow, Russia\\
}
\begin{document}
\maketitle
\begin{abstract}
{ The topological term in the effective action for the
electrically  neutral BCS system is discussed. It is applied for
the calculation of the transverse force acting on the vortex in
the limit of the smooth vortex core and vanishing interlevel
distance in the vortex core. The controversy between the
topological terms in cond-mat/9703124 and cond-mat/9411047 is
resolved.}
\end{abstract}
cond-mat/9703143
\vfill\eject
\section{Introduction}

There are many papers devoted to the effective action for the
BCS system. They differ by the topological terms: while in most of
papers the topological term is $n\dot\phi/2$, where $n$ is the
particle density and $\phi$ is the phase of the order parameter
(see eg \cite{Aitchison}), in the several other papers it is
$(n-C_0)\dot\phi/2$, where $C_0=p_F^3/3\pi^2$ is the particle
density in the normal metal (see eg \cite{Otterlo}). From these
different topological actions different predictions are made for
the transverse force, acting on the moving vortex. Here we discuss
this controversy.

Actually both actions give the correct hydrodynamic equations
for the BCS system in the uncharged limit. This is because the
difference between the topological actions is the total time
derivative if the parameter $C_0$ is treated as the dynamical
invariant. This would mean that from both actions one should come
to the same prediction for the vortex dynamics. However the latter
depends on the proper treatment of the time derivative of the
action. On the other hand, in the presence of the vortex, the phase
$\phi$ is not defined globally, which makes means that the
topological action in terms of $\phi$ is badly defined. That is why
we propose another form of the topological term effective action,
which does not contain the phase of the order parameter explicitly,
and which follows from the gradient expansion of the effective
action.

This action allows us to find the transverse force acting on
the vortex in the special limit case, when the bosonic effective
action provides the complete description of the vortex dynamics, ie
in the case when the quantization of the levels in the vortex core
can be neglected and the vortex core is smooth on the scale of the
coherence length. In this limit our result for the transverse force
coincides with that obtained in microscopic calculations
\cite{Kopnin1996}.

\section{Effective BCS action at
$T=0$.}

The calculated topological term in BCS action can be expressed
in terms of the order parameter (gap function) $\Delta$ in the
following form:
$$S_{top}={1\over 2i}\int d^3r \int_0^1d\tau \int_{-\infty}
^{\infty}dt~{\partial n\over \partial (\vert\Delta \vert^2)}
\left(
{\partial \Delta\over \partial t} {\partial \Delta^*\over
\partial\tau} -{\partial \Delta\over \partial \tau}
{\partial
\Delta^*\over
\partial t}\right)~~.\eqno(2.1)$$

Here $\tau$ is an auxiliary coordinate, it is introduced when
the effective action ${\bf Tr}\ln G^{-1}$ is presented as ${\bf
Tr}\int_0^1 d\tau G\partial_\tau G^{-1}$. The dependence on $\tau $
is chosen in such a way that at $\tau=0$ the coupling and the gap
function are absent, $\Delta(\vec r,t,\tau)\vert_{\tau=0}=0$,
while $\tau=1$ corresponds to the physical 3+1 dimensions, ie
$\Delta(\vec r,t,\tau)\vert_{\tau=1}= \Delta(\vec r,t)$.
In Eq.(2.1) $n$ is the particle density, which depends on $\Delta$.
At $\tau=0$ one has
$n(\tau=0)=p_F^3/3\pi^2\equiv C_0$, which is the particle number
density in the absence of the gap at the same chemical potential.

The quantity $n-C_0$ is small in the weak coupling limit of  the
BCS model, where it is determined by the particle-hole asymmetry,
$n-C_0\sim n\Delta^2/E_F^2 \ll n$. In a smooth crossover from the
BCS superconductivity to the condensate of the Cooper pairs,
discussed in many papers (see recent papers
\cite{Geshkenbein,Stintzing,Marini}), the parameter
$C_0$ decreases and can become  zero at some value of
the coupling parameter
$g$: $C_0(g)=0$ at $g>g_c$. In this case at $g=g_c$ one has the
quantum (Lifshitz) transition at $T=0$. This zero temperature
transition definitely happens if the quasiparticle spectrum has
nodes, eg in the case of the $d$-wave Cooper pairing or in the
$p$-wave state with the symmetry of the A-phase of superfluid
$^3$He. In this case at $g=g_c$ the spectrum of quasiparticles is
reconstructed -- the nodes disappear (see Sections 6.2 and 9.4 in
\cite{Exotic}). This is similar to the Lifshitz transition in
metals, where the topology of Fermi surface  changes. In what
follows we assume that we are on the BCS (weak coupling) side, ie
below the Lifshitz point,  $g<g_c$.

In the BCS action the relevant variable is the the gap function
$\Delta$,
which means that all other variables,
including $n$ are the functions of
$\vert\Delta
\vert^2$. The variation of the action over the order parameter
is the surface integral in the 5-dimensional space $({\bf
r},\tau,t)$ and thus is expressed in terms of the physical
coordinates $({\bf r},t)$:
$$\delta S_{top}={1\over 2i}\int d^3r  \int_{-\infty}
^{\infty}dt~{\partial n\over \partial (\vert\Delta \vert^2)}
\left(
{\partial \Delta\over \partial t}\delta \Delta^* -
\delta \Delta {\partial
\Delta^*\over
\partial t}\right)~~.\eqno(2.2a)$$
In principle from the variation of this topological action one can
restore the action in the physical space-time, but the well
defined action
contains explicitely the parameter
$C_0=n(\vert\Delta \vert^2=0)=p_F^3/3\pi$:
$$S_{top}={1\over 4i}\int d^3r  \int_{-\infty}
^{\infty}dt~( n-n(\vert\Delta \vert^2=0))
\left( {\dot \Delta^*\over
\Delta^*} -{\dot \Delta\over
\Delta}\right)~~.\eqno(2.2b)$$

If $\Delta({\bf r},t)$ is nowhere zero (this requirement excludes
the
case when vortices are present) one can introduce
the phase $\phi$ of the order parameter:
$$\Delta = \vert\Delta \vert e^{i\phi}~~.\eqno(2.3)$$
This allows to express the variation of the topological action in
Eq.(2.2a) in terms of the canonically conjugated variables, $n$ and
$\phi$:
$$\delta S_{top}={1\over 2}\int d^3r  \int_{-\infty}
^{\infty}dt~\left({\partial n\over \partial t}\delta \phi -
{\partial \phi\over \partial t}\delta n\right)~~,\eqno(2.4a)$$
where
$$\delta n={\partial n\over \partial (\vert\Delta \vert^2)}\delta
(\vert\Delta
\vert^2)~~.\eqno(2.5)$$
The action in Eq.(2.2b) can be also used in this case, it gives
$$S_{top}=-{1\over 2}\int d^3r  \int_{-\infty}
^{\infty}dt~(n-C_0) {\partial \phi\over \partial t}
~~.\eqno(2.4b)$$
This is what was obtained in \cite{Otterlo}.

The conventional contribution to the action is
$$S_0=\int d^3r  \int_{-\infty}
^{\infty}dt~\left({1\over 2}mn_s{\bf v}_s^2 +\epsilon( \vert\Delta
\vert^2)\right)~~,\eqno(2.6)$$
where $n_s$ is the superfluid density,
$m$ is the mass of the fermion, ${\bf
v}_s={\hbar\over 2m}\nabla\phi$ is the superfluid velocity and
$\epsilon$ is the energy density, which depends on $\vert\Delta
\vert^2$. Varying
$S=S_0+S_{top}$  over $\phi$ and
$\vert\Delta
\vert^2$ one obtains the conventional hydrodynamic equations (we
neglected the nonlinear terms in this procedure)
$${\partial n\over \partial t}+{\bf \nabla}\cdot(n_s{\bf
v}_s)=0~~,\eqno(2.7)$$
$${\partial \phi\over \partial t}=-2\mu~~,
~~\mu ={ \partial \epsilon/
\partial (\vert\Delta \vert^2) \over
 \partial n/ \partial (\vert\Delta
\vert^2)}~~.\eqno(2.8)$$

It is important here, that the hydrodynamic equations
are general and do not contain the  parameter $C_0$. This is
because the term containing the factor $C_0$ in Eq.(2.4b) is
the full derivative. In this sense there is no difference between
the topological terms discussed in \cite{Aitchison} and
\cite{Otterlo}. The difference becomes important when the zeroes in
$\Delta$ appear and the phase of the order parameter is not defined
globally any more. This is just the case of  vortices. To find the
vortex dynamics we must return to the action in Eq.(2.2a) for the
order parameter, which does not contain the phase
$\phi$ explicitly.

\section{Effective action for the vortex dynamics in continuous
limit.}

The  effective action  $S=S_0+S_{top}$  can be applied for the
derivation of the dynamics of the vortex line only under certain
conditions. We assume that the description in terms of the order
parameter is complete, ie there is no other degrees of freedom. All
the fermionic degrees of freedom are assumed to be inegrated off
when the effective action was obtained. In this integration the
fermionic spectrum was considered  classically, ie  the spectrum
was taken as a function of the commuting spatial coordinate ${\bf
r}$ and momentum ${\bf p}$:
$E=\sqrt{\varepsilon^2({\bf p})+\vert\Delta({\bf r})\vert^2}$.
This means that if one  applies this effective action to  vortices
one neglects the quantization of the fermions in the vortex
background. This is justified only when the distance between the
energy levels are small compared to the width of the levels,
$\omega_0\tau
\ll 1$. This represents the necessary condition for the application
of the effective action to the vortex motion.  Another important
condition is that the core size of the vortex is to be large
compared to the coherence length in order to neglect the higher
order gradient terms in the action.

Under these conditions we can show that the BCS effective action
leads  to the following motion equation for the vortex:
$$\pi N \hbar (n-C_0)\hat z\times {\bf v}_L=
\pi N\hbar n_s\hat
z\times{\bf v}_{s0} ~~.\eqno(3.1)$$
Here ${\bf v}_L$ is the vortex velocity with respect to the heat
bath (normal component or crystal lattice), which is here assumed
to be   at rest; ${\bf v}_{s0}$ is the superfluid velocity of the
external superflow; $N$ is the winding number of the vortex. This
coincides with the result of the microscopic calculations for the
electrically neutral case in the limit of the large core size and
in the regime
$\omega_0\tau \ll 1$ \cite{Kopnin1996}.

To get the Eq.(3.1) from the BCS action let us introduce the
vortex coordinate ${\bf r}_L(t)$. For simplicity we consider the
rectilinear vortex  along the axis $z$, so the order parameter
$\Delta$ depends only on ${\bf r}-{\bf r}_L(t)$ where both vectors
are 2-dimensional:
$$\Delta({\bf r},t)=\Delta({\bf
r}-{\bf
r}_L(t))~~.\eqno(3.2)$$
The variation of the topological action in
Eq.(2.2) becomes
$$\delta S_{top}={1\over 2i}\int d^3r ~{\partial n\over
\partial(\vert\Delta
\vert^2)}\left( {\partial \Delta\over \partial x}  {\partial
\Delta^*\over
\partial y} -{\partial \Delta\over \partial y}  {\partial
\Delta^*\over
\partial x}\right) \int_{-\infty}^{\infty}dt\left( {\partial
x_L\over \partial t} \delta y_L  - \delta x_L  {\partial
y_L\over
\partial t}\right)
~~.\eqno(3.3)$$

For the axisymmetric vortex with
$$\Delta(r,\varphi) =  a(r) e^{iN\varphi}~~,\eqno(3.4)$$
where $r$ and $\varphi$ are cylindrical coordinates, and $N$ is
the winding number, one obtains
$$ -i\int d^3r ~{\partial n\over \partial (\vert\Delta
\vert^2)}\left( {\partial \Delta\over \partial x}
{\partial
\Delta^*\over
\partial y} -{\partial \Delta\over \partial y}
{\partial \Delta^*\over
\partial x}\right)=$$
$$=2\pi NL\int_0^\infty dr {\partial n\over \partial r}=
2\pi NL(n(\infty)-n(0))=2\pi NL(n-C_0)
~~,\eqno(3.5)$$
where $L$ is the length of the straight vortex line.
We took into account
that on the vortex  axis  ($r=0$) the order parameter is zero,
$a(0)=0$, and therefore
$n(0)=C_0$.

The same result can be obtained for any type of the
vortex with the winding number $N$, because the volume integral can
be transformed to the surface integral far from the vortex core,
which is determined only by $N$ and $n(\vert\Delta(\infty)\vert^2)
- n(\vert\Delta\vert^2=0)=n-C_0$.
As a result
$$\delta S_{top}=\pi \hbar N L (n-C_0)
\int_{-\infty}^{\infty}dt\left(
{\partial x_L\over \partial t} \delta y_L  - \delta x_L  {\partial
y_L\over
\partial t}\right)~~.\eqno(3.6)$$
The variation of this topological action over $\delta{\bf r}_L$
gives the force acting on the vortex, which is represented by the
term on the lhs of Eq.(3.1).

Note that the expression for the
variation of Eq.(2.2a):
$${\delta S_{top}\over \delta{\bf r}_L} ={1\over 2i}\int d^3r
~{\partial n\over \partial (\vert\Delta \vert^2)}
\left(
{\partial \Delta\over \partial t}\nabla \Delta^* -
\nabla \Delta {\partial
\Delta^*\over
\partial t}\right)~~\eqno(3.7)$$
transforms to the Eq.(10) in \cite{Kopnin1996}
if one neglects the coordinate  dependence of $\partial
n/\partial (\vert\Delta \vert^2)$. The latter is justified only in
the weak coupling limit, $g\ll 1$, explored in
\cite{Kopnin1996}, while our approach is not limited
by the weak coupling assumption: we assume only that $g<g_c$.

In
\cite{KopninLopatin} the Eq.(10) of \cite{Kopnin1996} is generalized
to the case of the anisotropic pairing with the momentum dependent
gap function, $\Delta({\bf p},{\bf r})$, but still in the weak
coupling limit.
We can generalize this to the arbitrary coupling strength
introducing the momentum dependent particle distribution function
$$n({\bf p},{\bf r})={1\over 2}\left(1- {\varepsilon ({\bf
p})\over \sqrt{\varepsilon^2({\bf
p})+\vert\Delta({\bf p},{\bf r})\vert^2}}\right)~~, ~~n({\bf
r})=2\int{d^3p\over (2\pi)^3}n({\bf p},{\bf r})~.\eqno(3.8)$$
Then
$$\delta S_{top}=-i \int{d^3pd^3r\over (2\pi)^3}  \int_{-\infty}
^{\infty}dt\times$$
$$\times{\partial n({\bf p},{\bf r})\over \partial
(\vert\Delta ({\bf p},{\bf r})\vert^2)}
\left(
{\partial \Delta({\bf p},{\bf r})\over \partial t}\delta
\Delta^*({\bf p},{\bf r}) -
\delta \Delta({\bf p},{\bf r}) {\partial
\Delta^*({\bf p},{\bf r})\over
\partial t}\right)~~.\eqno(3.9)$$
Applying the same procedure as before one obtains the Eq.(3.6) for
any (smooth) structure of the vortex core and for arbitrary
anisotropic pairing state. I am indebted to N.B. Kopnin for this
remark \cite{KopninPrivate}.

The rhs term in Eq.(3.1) is obtained from the kinetic energy
term in $S_0$ in Eq.(2.6). In the presence of the external
superflow
${\bf v}_{s0}$ the
relevant term is
$${\bf v}_{s0}\cdot\int d^3r~n_s{\bf v}_{s}={\bf v}_{s0}\cdot {\bf
P}~~.\eqno(3.10)$$
The linear momentum ${\bf P}$ related to the vortex coordinate is
$$  {\bf P}=\pi \hbar N n_s L\hat z \times  {\bf
r}_L~~.\eqno(3.11)$$
The variation of ${\bf v}_{s0}\cdot {\bf
P}$ over ${\bf r}_L$  gives the second term in Eq.(3.1).

The equation (3.1) is not Galilean invariant. The Galilean
invariance is restored by introducing the velocity of the normal
component
${\bf v}_n$, which coincides with the velocity of crystal lattice
in the case of superconductors, or with the heat bath of the normal
excitations in the case of superfluids.
This equation can be represented
as the balance of three  forces acting on the vortex
\cite{KopninVolovik}:
$$\pi  \hbar N \hat z \times [n ({\bf v}_s-{\bf v}_L) + C_0({\bf
v}_L-{\bf v}_n) +
(n-n_s) ({\bf v}_n-{\bf v}_s)]=0~~.\eqno(3.12)$$
These are correspondingly the Magnus, the spectral flow and
the Iordanskii forces in the terminology of
Ref.\cite{KopninVolovik}.

In the form of Eq.(3.1) the force balance was also obtained in
\cite{Otterlo1995}, again in the limit $\omega_0\tau \ll 1$;
the first
discussion of the lhs term in Eq.(3.1) as originating from the
topological action was probably in
\cite{Volovik1986} (see the paragraph containing the Eq.(10) in
\cite{Volovik1986}).

\section{Discussion.}

In conclusion, the effective topological actions discussed in
\cite{Aitchison} and
\cite{Otterlo} are equivalent to each other in the particular
case when the phase of the condensate is defined globally, and thus
in the absence of vortices. In this case the two topological
actions differ by the total time derivative and thus lead to the
same hydrodynamic equations  for the conjugated variables, the
particle density  $n$ and the condensate phase $\phi$.  However
these actions cannot be applied when the zeroes in the order
parameter are present. In the presence of zeroes the topological
action in Eq.(2.2) (or in Eq.(3.9) for the more general case of
anisotropic pairing) is to be used, which resolves between the
discussed actions \cite{Aitchison} and
\cite{Otterlo} in favour of that in \cite{Otterlo}.

The Eq.(2.2) describes also the vortex dynamics in a special limit
case. This dynamics agrees with the general phenomenological
approach using the Poisson brackets scheme \cite{Volovik1996}. In
this approach the phenomenological parameter $C_0$ is introduced as
the dynamical invariant of the system, which does not violate
the general properties of the hydrodynamical Poisson brackets.
The BCS effective action for the neutral BCS system in Eq.(2.2)
gives the precise value of this phenomenological parameter,
$C_0=p_F^3/3\pi^2$, in the regime when the action is applicable for
the vortex dynamics. The conditions of the applicability are
$\omega_0\tau\ll 1$ and $R_{core}\gg \xi_0$, where $R_{core}$ is
the core radius. They correspond to the vortex with the smooth core.

In the opposite limit $\omega_0\tau=\infty$ the effective action
for the  vortex dynamics in the BCS system can also  be derived
from the first principles exploring the Berry phase (see the recent
paper \cite{Dziarmaga}). In this case the discrete spectrum of the
electrons in the core is of main importance. The properties of the
exact wave functions of the electron in the potential produced by
the vortex lead to the zero value of the canonical momentum density
at the origin, $S(0)=0$, in \cite{Dziarmaga}. This gives the zero
value of the parameter $C_0$ in Eq.(3.12) in complete agreement
with the result of \cite{KopninVolovik}, obtained in the limit
$\omega_0\tau=\infty$. The finite life time $\tau$ destroys the
coherence of the wave functions and finally restores the maximal
value of $C_0$ in the limit $\omega_0\tau\ll 1$.

I would like to thank Jacek Dziarmaga, Nikolai Kopnin and Anne van
Otterlo for stimulating discussions.


\begin{thebibliography}{15}

\bibitem{Aitchison} I.J.R. Aitchison, P. Ao,  D.J. Thouless and
X.-M. Zhu,
 Phys. Rev., {\bf B~51}, 6531 (1995).

\bibitem{Otterlo} A. van Otterlo, D.S. Golubev, A.D. Zaikin and G.
Blatter, cond-mat/9703124.

\bibitem{Kopnin1996} N.B.Kopnin, Phys.
Rev., {\bf B~54}, 9475 (1996).

\bibitem{Geshkenbein} V.B. Geshkenbein, L.B. Ioffe  and A.I.
Larkin, Phys. Rev., {\bf B~55}, 3173 (1997).

\bibitem{Stintzing} S. Stintzing and W. Zwerger, cond-mat/9703129.

\bibitem{Marini} M. Marini, F. Pistolesi and G.C.
Strinati, cond-mat/9703160.


\bibitem{Exotic} G. E. Volovik, {\it Exotic properties  of
superfluid $^3$He}, World Scientific, Singapore - New Jersey -
London - Hong Kong, 1992.

\bibitem{KopninLopatin} N.B.Kopnin and A.V. Lopatin, to be published
in  Phys. Rev.  {\bf B}.

\bibitem{KopninPrivate} N.B. Kopnin, private communications.

\bibitem{KopninVolovik} N.B.Kopnin,  G.E.Volovik and \"U.Parts,
Europhys. Lett. {\bf  32}, 651 (1995).

\bibitem{Otterlo1995}  A. van Otterlo,  M.V. Feigelman,  V.B.
Geshkenbein and G. Blatter,  Phys. Rev. Lett. {\bf  75}, 3736
(1995).

\bibitem{Volovik1986} G.E. Volovik, Pis'ma ZhETF {\bf 44}, 144
(1986) [JETP Letters, {\bf 44}, 185
(1986)].

\bibitem{Volovik1996} G.E. Volovik, Pis'ma ZhETF {\bf 64}, 794
(1996) ( cond-mat/9610157 ).

\bibitem{Dziarmaga} J. Dziarmaga, Phys. Rev., {\bf B~53}, 6572
(1996).

\end{thebibliography}
\end{document}